\documentclass[aps,prb,twocolumn,superscriptaddress,showpacs]{revtex4}
\usepackage{graphicx}
\usepackage{epstopdf}
\usepackage{mathptmx}
\begin{document}

%\preprint{}

%\title{Comparison of different Tight-binding and $\mathbf{k\cdot p}$-approaches to electronic properties of semiconductor quantum dots}
\title{A comparison of atomistic and continuum theoretical approaches to determine electronic properties of GaN/AlN quantum dots}

\author{Oliver Marquardt}
\email{marquardt@mpie.de}
\affiliation{Max-Planck Institut f\"ur Eisenforschung, Max-Planck-Stra\ss e 1, D-40237 D\"usseldorf, Germany}
\author{Daniel Mourad}
\affiliation{Institute for Theoretical Physics, University of
Bremen, D-28359 Bremen, Germany}
\author{Stefan Schulz}
\altaffiliation[Now at:]{ Tyndall National Institute, Lee Maltings,
Prospect Row, Cork, Ireland} \affiliation{Institute for Theoretical
Physics, University of Bremen, D-28359 Bremen, Germany}
\author{Tilmann Hickel}
\affiliation{Max-Planck Institut f\"ur Eisenforschung, Max-Planck-Stra\ss e 1, D-40237 D\"usseldorf, Germany}
\author{Gerd Czycholl}
\affiliation{Institute for Theoretical Physics, University of
Bremen, D-28359 Bremen, Germany}
\author{J\"org Neugebauer}
\affiliation{Max-Planck Institut f\"ur Eisenforschung, Max-Planck-Stra\ss e 1, D-40237 D\"usseldorf, Germany}

%\thanks{}


\date{\today}

\begin{abstract}
In this work we present a comparison of multiband
$\mathbf{k}\cdot\mathbf{p}$ models, the effective bond-orbital
approach, and an empirical tight-binding model to calculate the
electronic structure for the example of a truncated pyramidal
GaN/AlN self-assembled quantum dot with a zincblende
structure. For the system under consideration, we find a very good
agreement between the results of the microscopic models and the
8-band $\mathbf{k}\cdot\mathbf{p}$ formalism, in contrast to a
6+2-band $\mathbf{k}\cdot\mathbf{p}$-model, where conduction band
and valence band are assumed to be decoupled. This indicates a
surprisingly strong coupling between conduction and valence band
states for the wide band gap materials GaN and AlN. Special
attention is paid to the possible influence of the weak spin-orbit
coupling on the localized single-particle wave functions of the
investigated structure.
\end{abstract}

\pacs{71.15.-m, 73.21.La, 73.22.Dj}
\keywords{$\mathbf{k\cdot p}$ formalism, tight-binding, EBOM, quantum dots, electronic properties}

\maketitle

%\section{\label{intro}}
%\subsection{}
%\subsubsection{}

%-----------------------------------------------------------------------------
%----------------------------------Introduction-------------------------------
%-----------------------------------------------------------------------------
\section{Introduction\label{sec:intro}}

Nitride-based semiconductor nanostructures are promising
materials due to their potential application in opto-electronic  and
high-power/temperature electronic devices.~\cite{PoBo97} AlN, GaN
and InN and their ternary and quarternary alloys in principle allow
the emission of the whole spectrum of visible light from red to
ultraviolet. Within the past years, increasing research interest has
been on the investigation of GaN/AlN quantum dots (QDs) in order to
develop single electron transistors,~\cite{KaYa2001} ultraviolet
sources~\cite{HoAr2004} and detectors.~\cite{RaRo96}

Group III-nitrides can crystallize in the thermodynamically stable
configuration with a wurtzite crystal structure and in the
meta-stable modification with a zincblende
structure.~\cite{LaHe2004} The great majority of wurtzite GaN/AlN
QDs are grown along the polar $[0001]$ direction. These structures
exhibit large spontaneous and strain induced polarization. These
effects lead to a large internal electrostatic field, which is very
unique to III-nitride heterostructures and has a significant effect
on the electronic and optical properties of QDs. The magnitude of
the electrostatic built-in field has been estimated to be in the
order of MV/cm.~\cite{WiSi98,SiPe2003} Such fields spatially
separate the electrons and holes, which leads to a reduction of the
oscillator strength and enhanced radiative
lifetimes.~\cite{SiPe2003,FoBa2003,BaSc2007}

In contrast, in the cubic GaN/AlN QD structures, the
spontaneous polarization is absent due to the higher crystal
symmetry.~\cite{NoSt2008} Furthermore, experimental data indicate
that the piezoelectric contributions are
small.~\cite{GoEn2004} Therefore, GaN/AlN QDs with a
zincblende structure are expected to have advantages in
optoelectronic devices. Recently, there has been an increasing
interest in cubic GaN/AlN QDs due to the improvement of their growth
process.~\cite{MaAd2000,AdMa2001,DaFe2001,SiPe2003,GoEn2004,GaGe2005}
In order to understand the optical properties of cubic GaN/AlN QDs,
the investigation of the electronic structure of these structures is
of major importance. For instance, the excitation energies and wave
functions are crucial ingredients for
carrier-carrier~\cite{SeNi2006} and carrier-phonon~\cite{NiGa2005}
scattering in nitride QD structures.

Different approaches have been developed to calculate the electronic
structure of semiconductor QDs. These methods range from continuum
approaches such as effective mass~\cite{GrSt95,WoHa96,ShGa2003} and
$\mathbf{k}\cdot\mathbf{p}$~\cite{Pr1998,StGr98,AnORe2000,FoBa2003}
approximations to atomistic models, e.g.
tight-binding~\cite{SaAr2002,SaKo2003,ScCz2005} and pseudopotential
approaches~\cite{WaKi99,WaWi2000,BeZu2005}.

The number of available theoretical models for the calculation of
the electronic structure makes an evaluation of these methods with respect to the accuracy of the investigated
material properties necessary.

While in atomistic descriptions the computational effort grows with
the number of involved atoms, the accuracy of continuum models
decreases when the structure's characteristic dimensions reach the length scale of
the atomic bonds.
On the other hand, the continuum models are not limited
to a maximum size of the structure.
Previously, different $\mathbf{k}\cdot\mathbf{p}$-models have been compared
with an atomistic empirical pseudopotential method for different semiconductor
systems.~\cite{FuWa97, WaWi2000, LuLi2006} These investigations revealed various shortcomings
in the continuum models resulting from the lack of atomistic description or an
insufficient number of involved bands. However, a comparison of atomistic
and continuum models employing the same number of involved bands is essential in order
to determine the accuracy of computationally less demanding models.

In this work, we perform a careful comparison of various atomistic
and continuum methods, namely an empirical tight-binding
(ETBM) model, an effective bond-orbital model (EBOM) and two
variations of the $\mathbf{k}\cdot\mathbf{p}$ approach, to calculate
the electronic structure of GaN/AlN QDs with a zincblende structure.

Since we use for all methods an equivalent set of input parameters, the output of the various approaches can be compared directly. We focus our attention on the
differences in the electron and hole wave functions and the corresponding single-particle energies.
A comparison of optical properties such as excitonic absorption or emission spectra,
which can be obtained in the framework of a full configuration scheme~\cite{BaGa2004},
is beyond the scope of present work.
%These results serve as input for the calculation of optical properties
%like excitonic absorption and emission spectra in future investigations, i.e.
%in the framework of a full-configuration scheme.~\cite{BaGa2004}}

In this study special attention is paid to the influence of the
small spin-orbit coupling on the results, an effect which has been
commonly neglected in III-nitride QD
systems.~\cite{AnORe2000,AnORe2001,FoBa2003,SaAr2002,SaAr2003,BaSc2005,ScSc2006}
However, recent investigations on wurtzite III-nitride QDs
show that neglecting the spin-orbit coupling leads to artificial
degeneracies of hole states.~\cite{WiSc2006,ScSc2008} The influence
of spin-orbit coupling on the properties of nanostructures can be
expected to be strongly nonlinear in empirical approaches, as it enters both the common bulk
parameter set and the geometry related part of the Hamiltonian.

The influence of the conduction band - valence band
coupling in the 8-band $\mathbf{k}\cdot\mathbf{p}$-model will be
shown to have a surprisingly large effect on the electron binding energies despite the
fact that GaN is a wide band gap material.

This paper is organized as follows: In Sec.~\ref{sec:method}, we
introduce the applied methods and their underlying concepts and approximations.
Section~\ref{sec:model} is dedicated to the GaN/AlN QD
geometry. The following section deals with the electronic structure
of these systems. The influence of the spin-orbit coupling will be
discussed in detail in Sec.~\ref{sec:so}. In Sec.~\ref{sec:CbVb} we compare
the results of the 6- and the 8-band $\mathbf{k}\cdot\mathbf{p}$
model.

%-----------------------------------------------------------------------------
%-------------------------------------Method----------------------------------
%-----------------------------------------------------------------------------
\section{Applied methods\label{sec:method}}
While atomistic models of different sophistication approaching the
electronic properties of semiconductor nano\-structures lead to an
increasing computational effort with the number of involved atoms,
continuum models may produce strong deviations from results obtained
in atomistic simulations. These deviations are expected to increase
with decreasing characteristic dimensions of the structure. The aim
of this paper is to provide a comparison of complementary approaches
to the electronic structure of GaN/AlN QDs in a cubic structure.

Previous studies compared the $\mathbf{k}\cdot\mathbf{p}$ formalism
with highly accurate but computationally expensive
empirical pseudopotential calculations.~\cite{WoZu96, FuWa97}

In this study we chose two microscopic approaches with various levels of
approximation which have been constructed such that they reproduce
the band structure in the
for optical application relevant region around the $\Gamma$-point.
The investigated methods are (i) the empirical tight-binding method
(ETBM), (ii) the effective bond-orbital model (EBOM) and (iii) the
$\mathbf{k}\cdot\mathbf{p}$ formalism employing different numbers of
bands.
The choice of equivalent input parameters in the investigated methods allows
a direct comparison of differences resulting purely from
the different level of approximation.
While the ETBM is the most accurate model of the investigated ones, the EBOM and the
$\mathbf{k}\cdot\mathbf{p}$-formalism allow a straight-forward study of different
material parameters on the electronic structure
as the input parameter set is fixed for the investigated material system.
In this chapter, we will introduce the investigated methods used to compute the
electronic structure of the model system.

%----------------------------------Tight-binding------------------------------
\subsection{Empirical tight-binding model (ETBM)\label{sec:tb}}

The key assumption of the tight-binding method is that the overlap
of atomic orbitals decreases rapidly with the distance of their
corresponding atoms, i.e., only Hamilton matrix elements (TB
parameters) between neighboring atoms (typically up to the second
or third nearest neighbor shell) have to be included. For the polar
semiconductors GaN and AlN considered in this study, the upper
valence band is mainly formed by the p-orbitals of the anions and
the conduction band from the s-orbitals of the
cations.~\cite{Phillips73}
We therefore apply an
$s_cp^3_a$ TB model,~\cite{ScCz2005} where each anion is described
by the outer valence orbitals per spin direction: $p_x$,
$p_y$ and $p_z$. The cations are modeled by a single $s$ orbital per
spin direction. Overlap matrix elements up to the second nearest
neighbors are included in our TB model. Following
Ref.~\onlinecite{Chadi77}, the spin-orbit component of the bulk
Hamiltonian $H^{\mathrm{bulk}}$ couples only $p$-orbitals at the
same atom. By analytical diagonalization of the TB Hamiltonian
$H^{\text{bulk}}$ for special $\mathbf{k}$ directions the electronic
dispersion is obtained as a function of the TB parameters. Equations
for the TB parameters can now be deduced in terms of the
Kohn-Luttinger-parameters $(\gamma_1,\gamma_2,\gamma_3)$, the single
particle energy gap $E_g$, the effective electron-mass $m_e$ and the
spin-orbit splitting $\Delta_{\text{so}}$ at the $\Gamma$-point.
Doing so, one TB-parameter has to be determined self-consistently to
reproduce the $L$-point energy of the split-off band.
This parametrization has been verified to correctly
describe the band structure region around the $\Gamma$-point. Since
we are dealing with the electronic properties of a
nanostructure formed from a direct gap semiconductor material here, mainly
this part of the bulk band structure  is of importance. Furthermore, due to the
large energetic splitting ($> 1$ eV) between the zone center
($\Gamma$-point) and the side valleys ($L$-, $X$-point), the quantum
confinement will not introduce a mixing between these states.
Therefore, the region around the $\Gamma$-point is expected to
dominate the QD states.

%--------------------------------------EBOM-----------------------------------
\subsection{Effective bond-orbital model (EBOM)\label{sec:ebom}}
In the EBOM the TB orbitals are replaced
by  so called \textit{effective orbitals} located on the sites of the
underlying lattice, thus neglecting the atomic basis of the
material. With respect to the zincblende structure which is considered in this study,
the underlying symmetry of the original crystal structure is changed to that of an
fcc  lattice with effective orbitals
\begin{equation}
\left| \mathbf{R}, i, \sigma \right\rangle_{\mathrm{eff}}, \quad i = s,p_x,p_y,p_z, \quad \sigma=\uparrow,\downarrow,
\end{equation}
on each Bravais lattice site $\mathbf{R}$. We note that this
approximation gives rise to an artificial change of the symmetry
from $T_d$ (zincblende) to $O_h$ (fcc).

An advantage of the EBOM approach is that it allows to directly relate
the underlying TB-parameters with the corresponding $\mathbf{k}\cdot\mathbf{p}$-Hamiltonian.
A self-consistent fitting of the ETBM parameters to the bulk band structure is therefore not needed.

A first EBOM parametrization by Chang~\cite{Chang88} incorporated three-center overlap integrals. This parametrization
was restricted to coupling up to nearest neighbors only, so that solely the $\Gamma$-point energies could be fitted to the
set of $\mathbf{k} \cdot \mathbf{p}$-parameters.
In the present work, we use the parametrization of
Loehr~\cite{Loehr94}, which includes hopping up to second-nearest
neighbors  to fit the band-structure of the bulk material to
the above mentioned set of parameters.
This parametrization additionally allows for a fit to the $X$-point energies of
the conduction band ($X_{1c}$), the degenerate hh-/lh-band ($X_{5v}$) and the split-off
band ($X_{3v}$).
The degeneracy of the hh-/lh- band at $X$
is subsequently lifted by the incorporation of spin-orbit-coupling
into the TB Hamiltonian.

\begin{figure}[t]
%\psfrag{Gamma}[tc][tc]{$\Gamma$}
%   \psfrag{L}[tc][tc]{$L$}
%   \psfrag{X}[tc][tc]{$X$}
\includegraphics[width=0.9\columnwidth]{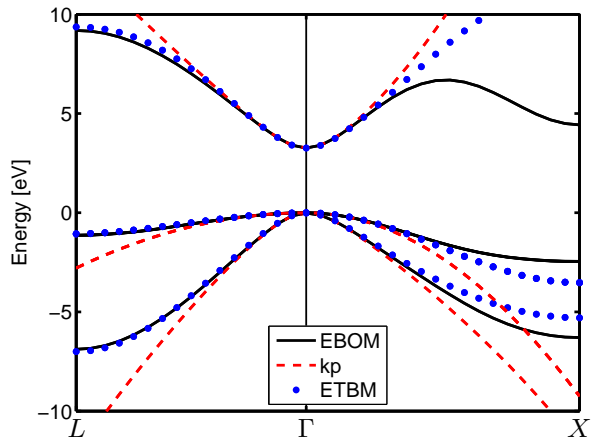}%{./bandstructureGaN.pdf}
\caption{(Color online) Bulk band structure of zincblende GaN in
two high symmetry directions calculated by the
effective-bond-orbital method (solid lines), the $s_cp^3_a$
tight-binding model (dotted lines) and the 8-band
$\mathbf{k}\cdot\mathbf{p}$ approach (dashed
lines).}\label{fig:compbandstruc}
\end{figure}

%--------------------------------------k.p------------------------------------
\subsection{$\mathbf{k\cdot p}$-method\label{sec:kp}}
In the $\mathbf{k}\cdot\mathbf{p}$-formalism the wave functions are
replaced by their envelope which no longer resolves individual
atoms. By this coarse graining treatment it becomes possible to
describe the electronic structure very efficiently on a continuum
scale. Within this work, we calculate the electronic structure in a
basis set of eight complex envelope eigenfunctions. This results in
an eight-band $\mathbf{k\cdot p}$ Hamiltonian~\cite{Bahder90}, given
in the Appendix. The eight-band formalism can be reduced to two plus
six bands, if the Kane parameter, $E_\mathrm{P}$, describing the
coupling between the valence band (VB) and the conduction band (CB)
is set to zero (see Appendix). Taking this parameter into account
causes modifications of the effective mass and Luttinger parameters
as well as additional coupling elements $U$ and $V$ within the
Hamiltonian matrix, $H^{8\times8}$, which are neglected in the
6+2-band model. Another common simplification which will be checked
explicitly in this study is to neglect the spin-orbit coupling
parameter $\Delta_\mathrm{so}$ which is in the order of a few
$\mathrm{meV}$ in the III-nitride systems.~\cite{VuMe2003} This
approximation reduces the dimensionality of the Hamiltonian and thus
the computational effort. The computation of the
$\mathbf{k}\cdot\mathbf{p}$ electron and hole wave functions is
performed in the plane-wave formalism within the S/PHI/nX software
package.\cite{sphinx,MaBoTBP}

%--------------------------------band structure-------------------------------
\subsection{Comparison of the bulk band structure\label{sec:bs}}

For the parametrization of the bulk band structure, we use
a set given by Fonoberov \emph{et al.}~\cite{FoBa2003},
which has been shown to accurately reproduce experimental data and
recent $\mathrm{G}_0\mathrm{W}_0$-calculations~\cite{RiWi2008}
around the $\Gamma$-point.

\begin{table}[t]%[H] add [H] placement to break table across pages
\caption{Material parameters for zincblende GaN and
AlN.~\cite{FoBa2003,FrSc2003}\label{tab:param}}
\begin{ruledtabular}
\begin{tabular}{ccc}
Parameters & GaN & AlN\\
\hline
$a$ [\AA] & 4.5 & 4.38\\
$E_\mathrm{g}$ [eV] & 3.26 & 4.9\\
$\Delta E_\mathrm{vb}$ [eV] & 0.8 & 0.0\\
$X_1^c$  [eV] & 4.428 &  5.346   \\
$X_3^v$  [eV] & -6.294 & -5.388  \\
$X_5^v$  [eV] & -2.459 & -2.315  \\
$E_\mathrm{P}$ [eV] & 25.0 & 27.1\\
$\Delta_{\mathrm{so}}$ [eV] & 0.017 & 0.019\\
$m_e$ [$m_0$] & 0.15 & 0.25\\
$\gamma_1$ & 2.67 & 1.92\\
$\gamma_2$ & 0.75 & 0.47\\
$\gamma_3$ & 1.10 & 0.85
\end{tabular}
\end{ruledtabular}
\end{table}
The parameter sets are given in
Tab. \ref{tab:param}. Figure~\ref{fig:compbandstruc} shows the
resulting band structure along the $L-\Gamma-X$ path for the three
methods. As can be seen, the agreement around the $\Gamma$-point is
perfect whereas expected clear deviations arise towards the
Brillouin-zone boundaries:
\\
The EBOM reproduces the bulk GaN valence-band structure obtained
from previous work~\cite{FrSc2003} throughout the Brillouin zone.
The ETBM results along the $\Gamma-L$ direction are in excellent
agreement with the EBOM valence-band structure, while slight
deviations from the EBOM results are observed at the $X$-point.
Remember that the EBOM energies have been fitted to this point of
the Brillouin zone.

Furthermore, the EBOM accurately describes the conduction band along
the $\Gamma-X$ direction. In particular, an additional maximum along
the $\Gamma-X$-direction is reproduced in agreement with ab-initio
band structure calculations. \cite{FrSc2003} The ETBM conduction band
deviates from the EBOM results near the X-point since higher
conduction-bands are not taken into account.

In contrast to the EBOM and ETBM, respectively, the 8-band
$\mathbf{k}\cdot\mathbf{p}$ reproduces the band structure only for
small $\mathbf{k}$-vectors around $\Gamma$. Similiar to the
discussion of the ETBM, by taking more bands into account a better
agreement of the $\mathbf{k}\cdot\mathbf{p}$ - band structure
throughout the Brillouin zone can be achieved.~\cite{StGo95}

However, in the present study we refrain from such an extended
$\mathbf{k}\cdot\mathbf{p}$ or TB Hamiltonian to keep the
number of involved bands equal and thus consistent within the investigated
models. Furthermore, in accordance with the discussion in
Sec.~\ref{sec:tb}, the $\Gamma$-point character is expected to
dominate the single particle states in a nanostructure with
characteristic dimensions of only a few $\mathrm{nm}$. Indeed,
as will be shown later, even the $8\times 8$ $\mathbf{k}\cdot\mathbf{p}$
Hamiltonian and the $s_cp^3_a$ TB model are excellent approximations.

%-----------------------------------------------------------------------------
%---------------------------------Results----------------------------------
%-----------------------------------------------------------------------------

\section{The model quantum dot\label{sec:model}}

Cubic GaN QDs embedded in AlN have been experimentally investigated
by various
groups.~\cite{SiPe2003,MaAd2000,AdMa2001,DaFe2001,GoEn2004,GaGe2005} These
studies have shown that such QDs grow
as truncated pyramidal structures.~\cite{MaAd2000,GoEn2004}
The QD's are commonly grown in a Stranski-Krastanov growth mode, i.e.,  they form
spontaneously when the wetting layer (WL) exceeds a critical thickness.

\begin{figure}
\includegraphics[width=.7\columnwidth]{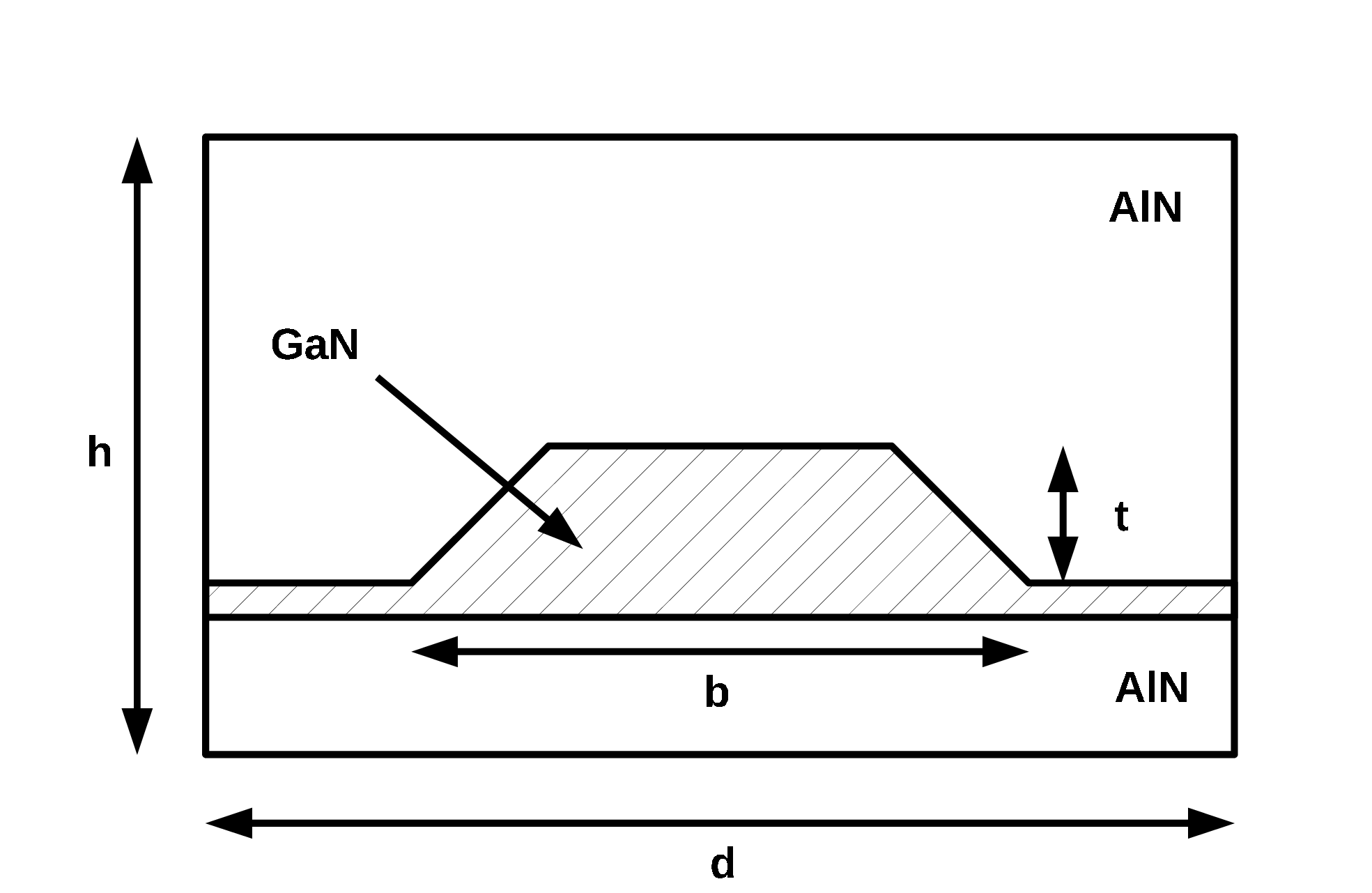}
\caption{Schematic geometry of our model quantum dot. The base length $b$ and the height $t$ of the pyramidal frustum determine the dot size, while $h$ and $d$ define the size of the supercell.
\label{fig:geom}}
\end{figure}

As pointed out before a major aim of this work
is to investigate the consistence of the results obtained by the continuum $\mathbf{k}\cdot\mathbf{p}$-approach with
the outcomes of (semi-) microscopical tight-binding approaches explained in Sec.~\ref{sec:method}.
%  As is known to be good for medium and large size structures also for a
While the accuracy of the continuum $\mathbf{k}\cdot\mathbf{p}$ model is known
to increase with the dimensions of the structure, a comparison between continuum and
atomistic simulations is highly interesting for systems with characteristic dimensions
in the order of magnitude of the bulk lattice constants. Besides, the small number of involved atoms
also limits the computational effort of the atomistic calculations.

As a test quantum dot we therefore consider a truncated pyramid with a square
base length of $b=16\,a$ and a height of $t=4\,a$ only.
It is placed on top of a GaN WL with a
thickness of $0.5a$, where $a$ is the AlN lattice constant. This
corresponds to $t\approx 1.75$ nm and $b=7$ nm. The dot is oriented along the
$\left[001\right]$  axis. Previous studies of
GaN/AlN QDs with a wurtzite structure showed no intermixing between
Ga and Al in these structures.~\cite{WiDa98,ArRo99} Since the
structural properties of cubic GaN QDs embedded in AlN are found to
be similar to those of the hexagonal ones~\cite{GoEn2004}, we also
take in our simulations the composition of the nanostructure and the
surrounding barrier material to be pure GaN and AlN, respectively.
Only for the WL composition experimentally a weak interdiffusion of
Ga and Al atoms is found.~\cite{ArRo99} Since the aim of this study
is on the bound single particle states, which are localized inside
the nanostructure, compositional fluctuations in the WL region have
only minor influence on the bound single particle states and will be
neglected. Note that the symmetry of the outer shape of the QD resulting from
the confinement potential is a $C_{4v}$-symmetry, while the underlying crystal
lattice lacks inversion symmetry and reduces the symmetry to
$C_{2v}$.

Within all three approaches, the QD is located in a sufficiently
large supercell to eliminate the influence of the chosen supercell
boundaries on the single particle states. The convergence of the
eigenstates with respect to the supercell size has been
carefully checked.

In the framework of a $s_cp_a^3$ TB model, the $C_{2v}$ symmetry of the QD's
underlying zincblende structure is naturally included. To set up the
Hamiltonian we use the TB parameters of the corresponding bulk
materials. At the GaN/AlN interfaces we use a linear interpolation
of the TB parameters of GaN and AlN. Since the nitrogen atoms form a
common anion lattice the interpolation affects only second nearest
neighbor elements which are small compared to the nearest
neighbor contributions.

The application of EBOM to describe QDs is similar to the ETBM
approach. The main difference is the restriction to a slightly more
coarse grained grid where the anion and cation positions cannot be resolved.
This coarse graining changes the underlying lattice from
zincblende to fcc. Therefore, this approach cannot sustain the
original $C_{2v}$ symmetry of the zincblende structure and increases
the number of symmetry operations. In this case we are left with a
$C_{4v}$ symmetry.

Within the $\mathbf{k\cdot p}$-formalism, the dot and the WL are
described by a spatially resolved envelope function. We use the bulk
GaN parameter set inside the nanostructure and bulk AlN parameters
for the matrix material. The $\mathbf{k\cdot p}$-formalism does not
resolve individual atoms and will therefore not reproduce the
$C_{2v}$ symmetry of the underlying zincblende lattice. The QD is
simulated on a real-space mesh of $80\times 80\times 80$ mesh
points.

In this study the focus is on a systematic comparison of the
different approaches introduced in Sec.~\ref{sec:method} rather than
on a complete description of all aspects of a QD. Therefore, we do
not consider contributions from strain and electrostatic built-in
fields in our calculations. Nevertheless, as we will discuss in the
following section, our results for the single-particle level
structure are in qualitative agreement with results obtained in
Ref.~\onlinecite{FoBa2003} where strain effects and
piezoelectric fields have been explicitly taken into account.

\section{Electronic properties of cubic $\mathbf{GaN}$ quantum
dots\label{sec:results}}
%------------------------------------states-----------------------------------
\subsection{Single particle energies and states\label{sec:states}}
\begin{figure}[t]
    \includegraphics[width=0.9\columnwidth]{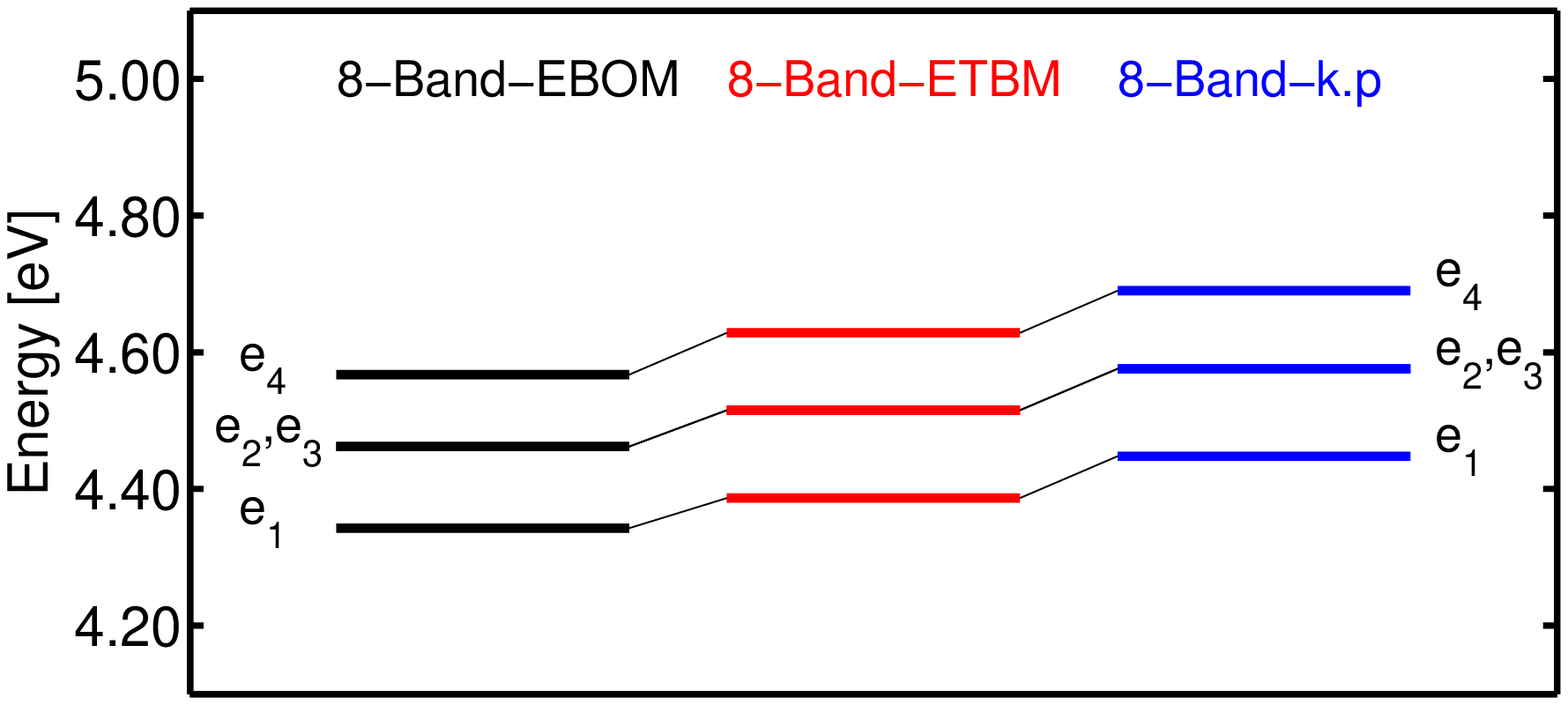}\\
    \hspace{9mm}$\sim$ \\
    \includegraphics[width=0.9\columnwidth]{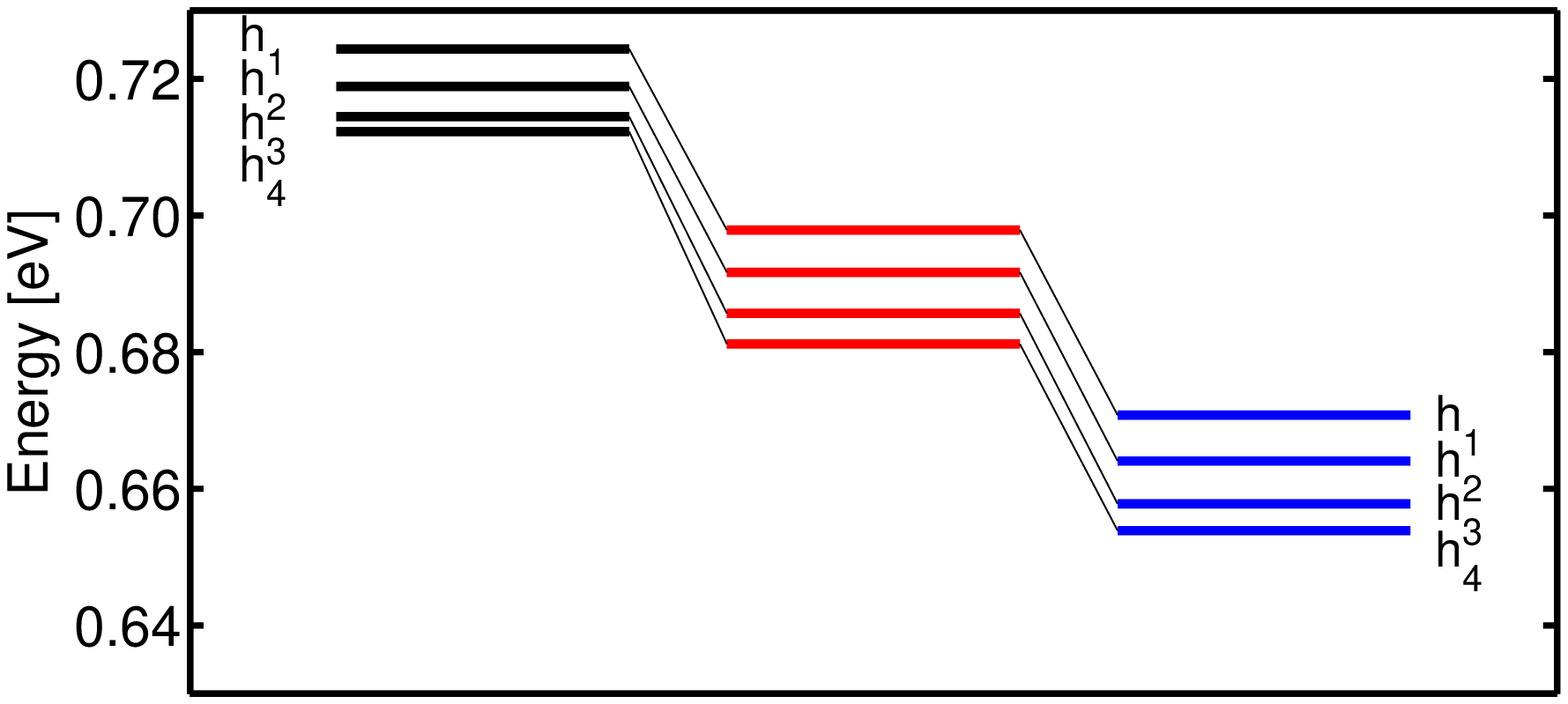}
    \caption{(Color online) The first four electron and hole single-particle
    energy levels of the GaN QD, as calculated by the effective-bond-orbital
    method (EBOM), the $s_cp^3_a$ empirical tight-binding model (ETBM) and the 8-band-
    $\mathbf{k} \cdot \mathbf{p}$ approach.
    All energies are given with respect to the valence band edge of the
    AlN. Please note the different scaling for the electron and hole energies, respectively.
    }\label{fig:singlepartenergies}
\end{figure}

Using the above quantum dot geometry, we applied the three approaches to compute the
bound hole and electron states. In all three approaches, we consistently find a total number
of eight bound electron states. Due to the large effective mass of the hole states and the
large valence-band offset the number of localized hole states
is much higher. While higher excited states play an important role in carrier-carrier
and carrier-phonon interaction, we will restrict our discussion on the first four
bound electron and hole states since these dominate the excitonic and absorption processes in
QD structures. The energy levels of these states as calculated by the three approaches are
shown in Fig.~\ref{fig:singlepartenergies}.

\begin{figure}%[t]
\includegraphics{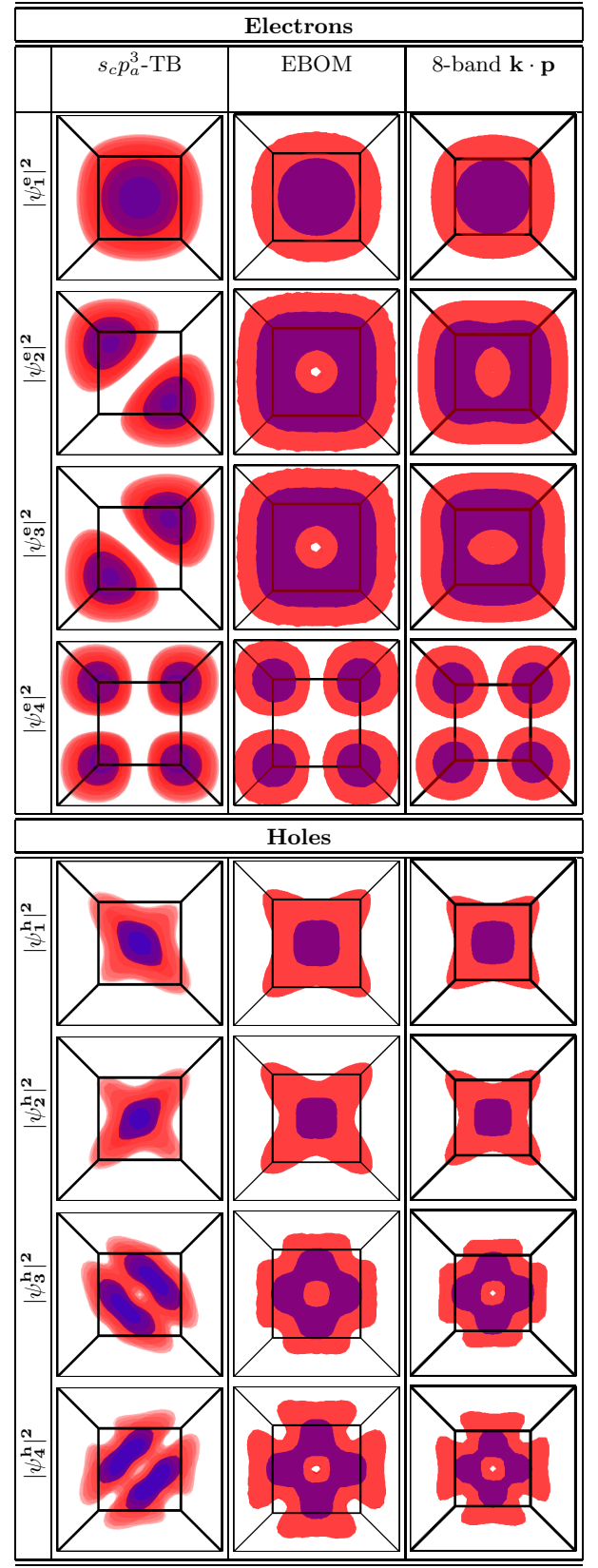}
\caption{(Color online) Top view of the truncated pyramidal GaN QD
structure with the first four bound states for electrons (upper
part) and holes (lower part). Depicted are isosurfaces of the
probability density with 10\% (red) and 50\% (blue) of the maximum
value.} \label{fig:oneparticle}
\end{figure}

%Figure~\ref{fig:singlepartenergies} shows the first four bound
%electron and hole single-particle energy levels, as calculated by
%the three above mentioned approaches.
%\emph{We focus our discussion on these states, since they are dominating
%the excitonic and absorption processes in QD structures.
%The total number of bound states is important
%if one is interested in carrier-carrier or carrier-phonon scattering processes.
%In all three approaches,
%we consistently find a total number of eight bound electron states.
%Due to the large effective mass of the hole states and the large valence-band
%offset the number of localized hole states is much higher.
%}
Fig.~\ref{fig:oneparticle} shows a top view of the QD geometry and
the modulus square $| \psi(\mathbf{r}) |^2$ of the first four
single-particle wave functions for electrons and holes,
respectively. Each state is twofold degenerate due to time reversal
symmetry. Qualitatively comparable results have been found by
Fonoberov \emph{et al.} for larger truncated pyramidal GaN/AlN
zincblende QDs.~\cite{FoBa2003}

\emph{Electron states:} Comparing the three approaches, the
single-particle states for the electrons are quantitatively as well as qualitatively very similar: Both their
energy eigenvalues and the corresponding symmetry character
agree well. According to their nodal structure, these states can be classified as $s$-,
$p_\mathrm{x}$-, $p_\mathrm{y}$- and $d$-like states, respectively. The electron ground state $\psi^{e}_1$ is
$s$-like, while the next two states are $p$-like. In case of the
$\mathbf{k}\cdot\mathbf{p}$ and the EBOM approach the symmetry of
the system is $C_{4v}$. Here, the atomic structure of the underlying
zincblende lattice is not resolved, and the states $\psi^e_2$ and
$\psi^e_3$ are energetically degenerate and form linear combinations of the form
$p_{\pm}=(1/\sqrt{2})(p_x+ip_y)$.
However, if taking the crystal structure into account, as it is
done in an empirical TB model or a pseudopotential approach, the
symmetry is reduced and degeneracies are lifted. For the QD considered here, a
truncated pyramidal GaN QD grown along the
$[001]$ direction and with  zincblende structure, the symmetry is $C_{2v}$.
This symmetry lifts the equivalence between the $[110]$ and $[1\bar{1}0]$ direction.
Employing the empirical TB model, we therefore obtain energetically non-degenerate
 $p_x$- and $p_y$-like states for $\psi^{e}_2$ and $\psi^{e}_3$,
respectively.  The states $\psi^{e}_2$ and $\psi^{e}_3$
are found to be non-degenerate with an energy difference of about $0.2$ meV.

This value is much smaller than the energy differences between the
lower bound states (see Figure~\ref{fig:singlepartenergies}). This
splitting may become more pronounced in other material systems or
for other QD geometries.~\cite{BeZu2005} Furthermore, inclusion of
an atomistic strain field and the piezoelectric potential may
also increase this splitting.~\cite{BeZu2005}

%\clearpage

\emph{Hole states:} In contrast to the electron states, the hole
states cannot be easily classified according to their nodal
structure. This is due to the strong intermixing of the various
valence bands and prevents a strict classification
of the optical selection rules on total angular momentum selection rules.
This finding emphasizes the importance of a multiband
approach. Qualitatively, we find an excellent agreement of the first
four hole states, and again the corresponding eigenvalues for all
three approaches lie within a narrow energy range. However, in case
of the 8-band $\mathbf{k}\cdot\mathbf{p}$ approach and the EBOM, the
first two hole states $\psi^{h}_1$ and $\psi^{h}_2$ reveal no
spatial anisotropy along the $[110]$ and $[1\bar{1}0]$ direction,
respectively. In the ETBM these states show a strong anisotropy
along the $[110]$ and $[1\bar{1}0]$ direction. This behavior again
reflects that only the ETBM approach correctly reproduces the
$C_{2v}$ symmetry of the system. Note that the first two hole
states  are not degenerate, and all three approaches yield a
splitting of about $6$ meV. This effect results (mainly) from the
spin-orbit splitting energy $\Delta_\mathrm{so}$ and will be
discussed in more detail in Sec.~\ref{sec:so}.

The splitting of the first two hole states and the anisotropy of
these states also has a strong influence on the optical properties
of these systems. For example, the splitting of the states
$\psi^{h}_1$ and $\psi^{h}_2$ may lead to additional lines in the
optical spectra. Furthermore, the absence of spatial anisotropy of
the states $\psi^{h}_1$ and $\psi^{h}_2$ may also lead to a
vanishing polarization anisotropy, $\lambda$, for dipole transitions
along directions $[110]$ and $[1\bar{1}0]$, respectively.
Energy differences in the absolute eigenvalues of electron and hole states
occur due to the different representation of the structure within the
investigated models, e.g, within the $\mathbf{k}\cdot\mathbf{p}$ formalism
the mesh discretization cannot resolve the microscopic representation (i.e. the
atomistic nature of the interface) of the
other two methods. Shifting the QD boundaries slightly by $\pm 3.5$ \r{A}
in the $\mathbf{k}\cdot\mathbf{p}$ formalism modifies the
absolute energies of electron and hole states about $\pm$ 15 meV but causes
no significant deviations in the energy difference of the states with respect to the
corresponding ground state.
Since the
electron and hole level structure obtained here from the different approaches is found
to be similar to the ones in Ref.~\onlinecite{BaSc2007} for a wurtzite InN/GaN QD,
similar excitonic and emission spectra are expected for the present system.
Selection rules for optical transitions can be analyzed in an analogous manner
from symmetry aspects, as for example discussed in Ref.~\onlinecite{BaSc2007}.

%----------------------------------spin-orbit---------------------------------
\begin{table}
\caption{Single-particle energies for the truncated pyramidal GaN QD
with ($\Delta_{\mathrm{so}}\neq 0$) and without
($\Delta_{\mathrm{so}}=0$) spin-orbit coupling. Each of the given
states is two-fold degenerate due to spin and time reversal
symmetry, respectively.}
\label{tab:energiesSO}
\begin{tabular}{|c|c|c|c|c|}%c|c|c|c|}
\hline
 & \multicolumn{4}{c|}{$\Delta_{\mathrm{so}}=0$} \\ \hline
 & 3+1-band $\mathbf{k}\cdot\mathbf{p}$ & 4-band $\mathbf{k}\cdot\mathbf{p}$  & EBOM & ETBM\\ \hline
 $e_1$ [eV] & 4.5259 & 4.4477 & 4.3428 & 4.4246 \\
 $e_2$ [eV] & 4.6768 & 4.5759 & 4.4621 & 4.5677 \\
 $e_3$ [eV] & 4.6768 & 4.5759 & 4.4621 & 4.5677 \\
 $e_4$ [eV] & 4.8069 & 4.6897 & 4.5670 & 4.6944 \\ \hline
 $h_1$ [eV] & 0.6679 & 0.6726 & 0.7264 & 0.7070 \\
 $h_2$ [eV] & 0.6679 & 0.6726 & 0.7264 & 0.7063 \\
 $h_3$ [eV] & 0.6622 & 0.6627 & 0.7193 & 0.6962 \\
 $h_4$ [eV] & 0.6558 & 0.6591 & 0.7172 & 0.6906 \\ \hline
 & \multicolumn{4}{c|}{$\Delta_{\mathrm{so}}=17~\mathrm{meV}$} \\ \hline
 & 6+2-band $\mathbf{k}\cdot\mathbf{p}$ & 8-band $\mathbf{k}\cdot\mathbf{p}$ & EBOM & ETBM\\ \hline
 $e_1$ [eV] & 4.5259 & 4.4479 & 4.3429 & 4.3866 \\
 $e_2$ [eV] & 4.6768 & 4.5761 & 4.4623 & 4.5152 \\
 $e_3$ [eV] & 4.6768 & 4.5761 & 4.4623 & 4.5154 \\
 $e_4$ [eV] & 4.8069 & 4.6900 & 4.5672 & 4.6284 \\ \hline
 $h_1$ [eV] & 0.6677 & 0.6708 & 0.7244 & 0.6979 \\
 $h_2$ [eV] & 0.6614 & 0.6641 & 0.7189 & 0.6917 \\
 $h_3$ [eV] & 0.6570 & 0.6578 & 0.7145 & 0.6857 \\
 $h_4$ [eV] & 0.6505 & 0.6539 & 0.7123 & 0.6812 \\ \hline
\end{tabular}
\end{table}

\subsection{Influence of Spin-Orbit-Splitting\label{sec:so}}

The spin-orbit coupling has been commonly neglected in previous
studies of nitride based
nanostructures,~\cite{AnORe2000,AnORe2001,FoBa2003,SaAr2002,SaAr2003,BaSc2005,ScSc2006}
since in group III-nitrides this contribution is of the order of a
few meV.~\cite{VuMe2003} The spin-orbit splitting is a relativistic
effect which increases with the atomic number of the
atoms.~\cite{YuCa2001} For example for CdSe and Znse this splitting
is in the order of 0.4 eV.~\cite{ScCz2005} Previous ETBM studies for
CdSe/ZnSe QDs comparable in shape and size reveal a splitting
in the first two hole states of several
meV.~\cite{ScCz2005,ScSc2007} Without spin-orbit coupling these
states are degenerate in the framework of a continuum approach,
similar to the results reported in Ref.~\onlinecite{FoBa2003} for a
truncated pyramidal GaN QD with a zincblende structure. Recently,
the influence of the spin-orbit coupling was shown to break the
degeneracy of the first two hole states in wurtzite InN/GaN
QDs.~\cite{WiSc2006,ScSc2008} In this section we will therefore
discuss the influence of the spin-orbit coupling on the electronic
properties.

\begin{figure}[t]
\includegraphics[width=0.9\columnwidth]{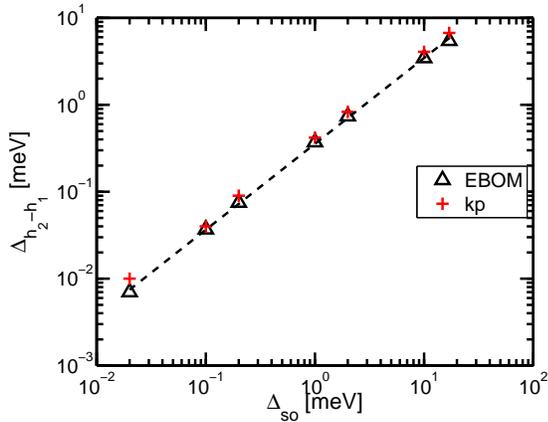}
\caption{(Color online) Energy difference $\Delta_{h_2-h_1}$ between
the first  two hole states in dependence of artificially varied
spin-orbit coupling as calculated with the EBOM and the
$\mathbf{k}\cdot\mathbf{p}$ model. The parameter
$\Delta_{\mathrm{so}}$ goes over several orders of magnitude,
note the logarithmic scale, from 0 to an upper value given by
 $\Delta_{\mathrm{so}}=17$ meV for GaN. The level splitting depends almost linearly on the bulk spin-orbit-splitting.}\label{fig:splitting}
\end{figure}
%
%
%
%Results 8-Band/EBOM/TB without SO coupling
%

In Tab. \ref{tab:energiesSO}  the first four electron and hole
energies for all three models are given in the presence as well as
in the absence of spin-orbit coupling. Interestingly, the essential
Kramers degeneracy resulting in twofold degenerate states
left aside, all calculations in the present paper bear a lift of
degeneracy between the first two hole states of about 5-6 meV for
the GaN dots, thus in the same order of magnitude as in the CdSe
dots when spin-orbit coupling is included.~\cite{ScCz2005}

Since the EBOM
parameters are not
updated in a self-consistent manner, these models provide the
opportunity to study the influence of the SO-coupling
$\Delta_{\mathrm{so}}$ on the single-particle states and energies by
gradually increasing this parameter from zero (which gives the limit
of a four-band model) up to the final value of 17 meV in GaN and 19
meV in AlN, respectively.

The calculated dependence of the level splitting $\Delta_{h_2-h_1}$
as function of $\Delta_{\mathrm{so}}$ is given in Fig.
\ref{fig:splitting}. It is linear over the entire range. As
the bulk band structures of GaN and AlN are hardly altered by the
relatively small spin-orbit coupling, the main influence on the
energy levels stems from the site-diagonal incorporation of the
spin-orbit coupling into the nanostructure Hamiltonian. Additional
$\mathbf{k}\cdot\mathbf{p}$-calculations give comparable results
which are also depicted in Fig. \ref{fig:splitting}.

Hence, we have demonstrated that for the system under consideration,
despite its relatively small influence on the bulk band structure,
the inclusion of SO-coupling allows to lift the artificial
degeneracy of hole states and therefore is of essential importance
for an accurate description of the single particle states.

%----------------------------------6-and 8-Band-Results---------------------------------
\subsection{Influence of CB-VB coupling\label{sec:CbVb}}
While the eight-band formalism yields good agreement with the
(semi-) microscopic EBOM and ETBM methods, a six-band approach combined with
an effective mass model provides reliable information for the hole
states only. Previous studies\cite{PoFo2001,Pr1998} found a strong
influence of conduction band-valence band-coupling for small and
medium band gap materials. Despite the fact that GaN has a large
band gap of $3.26$ eV, an unexpectedly large coupling is also
observed for the system considered here: coupling effects between
the conduction and valence bands significantly modify the electron
binding energies. The corresponding values are given in Tab.
\ref{tab:energiesSO}. However, these couplings have essentially
no effect on the nodal character of the wave functions. Comparing
the difference of the $2^{\mathrm{nd}}$ and $3^{\mathrm{rd}}$
electron state binding energy to the ground state energy, we find
$0.1282$ eV applying the eight-band model, which is in excellent
agreement with results from ETBM ($0.1286$ and $0.1288$ eV) and EBOM
($0.1194$ eV) calculations. The 6+2-band $\mathbf{k}\cdot\mathbf{p}$
approach which is for the electrons essentially an effective mass
approach, gives an energy difference of $0.1509$ eV, i.e., it
overestimates these energies by about $23$ meV ($17\%$). The origin of the
rather large deviation is due to the value of the Kane matrix
parameter, $E_\mathrm{P}$, which describes the CB-VB mixing effects
and contains  the respective dipole matrix element: Its value is
large enough ($E_\mathrm{P}$ is $25$ eV in GaN and $27.1$ eV in AlN)
to produce non-negligible coupling effects even for these
wide band gap materials. This emphasizes the fact, that not
only the energy gap is important for a possible decoupling of the
conduction- and valence-bands, but also the magnitude of the Kane
parameter is of crucial importance.

Neglecting the spin-orbit coupling in the 6+2-band approach leads to a 3+1-band
model. The energies are given in Tab. \ref{tab:energiesSO}. While the electron
binding energies remain
unchanged, we find a similar behavior for the hole states as we find using the
4-band model by neglecting $\Delta_{\mathrm{so}}$ in the 8-band model: the first
two hole states are found
to be degenerate in this model. When including the spin-orbit coupling, the resulting energy difference between the $1^{\mathrm{st}}$ and the $2^{\mathrm{nd}}$ hole state
is $6.3$ meV in the 6+2-band and $6.7$ meV within the full 8-band approach.
For the hole states, only small differences between the eight- and the six-band model are found. Again, we find that the nodal character for the hole state wave functions as found by the eight-band model is preserved when neglecting $E_\mathrm{P}$.

%-----------------------------------summary-----------------------------------
\section{Summary and discussion\label{sec:summary}}

In conclusion, we applied atomistic and continuum models, namely $\mathbf{k}\cdot\mathbf{p}$-models of different levels of sophistication,
an EBOM and an $s_\mathrm{c}p_\mathrm{a}^3$ ETBM model to derive the electronic properties of a zincblende GaN/AlN quantum dot.
Starting from a set of equivalent parameters fitted to the bulk band structure around the $\Gamma$-point for all methods, and applying
them to the same QD model structure,
we find a satisfactory agreement between the investigated models for the electron and hole wave functions and binding energies.
This demonstrates that for the GaN/AlN system also the semi-microscopic EBOM and
the continuum 8-band $\mathbf{k}\cdot\mathbf{p}$ model are appropiate
to describe the electronic properties of nanostructures down to feature sizes of a
few nm.
Small discrepancies between the ETBM on the one side and the EBOM and 8-band $\mathbf{k}\cdot\mathbf{p}$ models on the other side
are found to be a result of the underlying crystal symmetry described correctly in the ETBM only.\\
Despite the large band gap of GaN and AlN, we find strong deviations
of the electron binding energies between the 8-band model and a
decoupled 6+2-band approach. These occur due to the strong influence
of the Kane parameter $E_\mathrm{P}$ even in wide band gap
materials.

The commonly neglected spin-orbit splitting parameter
$\Delta_\mathrm{so}$ lifts the degeneracy of the first two hole
states. Even though $\Delta_\mathrm{so}$ is much smaller in GaN and
AlN than in e.g. CdSe, the resulting splitting is in the same order
of magnitude in both material systems. Neglecting this parameter is
therefore not suitable in the studied material system. By artificially varying
$\Delta_\mathrm{so}$, we find a strongly linear
correlation between the spin-orbit splitting and the energy
difference between the first two hole states.

\begin{acknowledgments}
The authors would like to thank Paul Gartner, Liverios Lymperakis,
Patrick Rinke and Frank Jahnke for fruitful discussions as well as
Sixten Boeck and Christoph Freysoldt for support concerning the
S/Phi/nX library. This work has been supported by the Deutsche
Forschungsgemeinschaft (Research group ``Physics of nitride-based,
nanostructured, light-emitting devices'', project Ne 428/6-3 and
project Cz 31/14-3). Stefan Schulz was further supported by the
Humboldt Foundation through a Feodor-Lynen research fellowship.
Daniel Mourad, Stefan Schulz and Gerd Czycholl also acknowledge a
grant for CPU time from the NIC at the Forschungszentrum J\"ulich.
\end{acknowledgments}
\newpage
\appendix

\section{8-Band $\mathbf{k}\cdot\mathbf{p}$ Hamiltonian\label{appA}} %{Appendix A\label{appA}}
%\textbf{The 8-Band $\mathbf{k}\cdot\mathbf{p}$-formalism}
In a basis set of eight complex wave functions
$$\Psi =
\left(\psi^{\Gamma_6}_{-\frac{1}{2}},
\psi^{\Gamma_6}_{\frac{1}{2}},\psi^{\Gamma_8}_{-\frac{3}{2}},
\psi^{\Gamma_8}_{-\frac{1}{2}} ,\psi^{\Gamma_8}_{\frac{1}{2}},
\psi^{\Gamma_8}_{\frac{3}{2}},\psi^{\Gamma_7}_{-\frac{1}{2}},
\psi^{\Gamma_7}_{-\frac{1}{2}}\right)$$
where $\Gamma_6$  denotes the conduction-, $\Gamma_8$ the light and heavy hole valence band states and $\Gamma_7$ denotes the spin-orbit coupling,
the eight-band $\mathbf{k}\cdot\mathbf{p}$ Hamiltonian~\cite{Bahder90} can be written as:
\begin{widetext}
\begin{equation}
\hat{H}^{8\times 8}=\left(\begin{array}{cc}
\hat{H}_{c} & \hat{H}_{s}\\
\hat{H}_{s}^\star & \hat{H}_{v}
\end{array}\right) =
\left(\begin{array}{cccccccc}
A & 0 & V^\star & 0 & \sqrt{3}V & -\sqrt{2}U & -U & \sqrt{2}V^\star\\
0 & A& -\sqrt{2}U & -\sqrt{3}V^\star & 0 & -V & \sqrt{2}V & U\\
V & -\sqrt{2}U &-P+Q & -S^\star & R &0 &\sqrt{\frac{3}{2}}S & -\sqrt{2}Q\\
0 & -\sqrt{3}V & -S & -P-Q & 0 & R & -\sqrt{2}R & \frac{1}{\sqrt{2}}S\\
\sqrt{3}V^\star & 0 & R^\star & 0 & -P-Q & S^\star & \frac{1}{\sqrt{2}}S^\star & \sqrt{2}R^\star\\
-\sqrt{2}U & -V^\star & 0 & R^\star & S & -P+Q & \sqrt{2}Q & \sqrt{\frac{3}{2}}S^\star\\
-U & \sqrt{2}V^\star & \sqrt{\frac{3}{2}}S^\star & -\sqrt{2}R^\star & \frac{1}{\sqrt{2}}S & \sqrt{2} Q & -P-\Delta_{\mathrm{so}} & 0\\
\sqrt{2}V & U & -\sqrt{2}Q & \frac{1}{\sqrt{2}}S^\star & \sqrt{2}R & \sqrt{\frac{3}{2}} S & 0 & -P-\Delta_{\mathrm{so}}
\end{array}\right)
\end{equation}
\end{widetext}
where the effective mass and the six-band model can be found in the $2\times 2$ $\hat{H}_{c}$
for the electron and the $6\times 6$ $\hat{H}_v$ for the hole states.
$\hat{H}_s$ denotes the superposition of electron and hole states within the eight-band model.
The matrix elements are given as:
\begin{eqnarray}
\label{eMass}
\nonumber A &=& E_\mathrm{cb}-\frac{\hbar^2}{2m_0}\gamma_c
\left(\partial^2_x + \partial^2_y + \partial^2_z\right)\\
%\displaystyle
\nonumber
P &=& -E_\mathrm{vb}-\gamma_1\frac{\hbar^2}{2m_0}\left(\partial^2_x + \partial^2_y + \partial^2_z\right)\\
%\displaystyle
\nonumber
Q &=& -\gamma_2\frac{\hbar^2}{2m_0}\left(\partial^2_x + \partial^2_y - 2\partial^2_z\right)\\
%\nonumber
R &=& \sqrt{3}\frac{\hbar^2}{2m_0}\left[\gamma_2\left(\partial^2_x - \partial^2_y\right) -2i\gamma_3\partial_x\partial_y\right],\\
\nonumber
S &=& -\sqrt{3}\gamma_3\frac{\hbar^2}{2m_0}\partial_z\left(\partial_x -i\partial_y\right),\\
\nonumber
U &=& \frac{-i}{\sqrt{3}}P_0\partial_z,\\
\nonumber
V &=& \frac{-i}{\sqrt{6}}P_0\left(\partial_x -
i\partial_y\right)
\end{eqnarray}
Note the minus sign appearing in the $S$ instead of the $R$ element in contrast to Ref. \onlinecite{Bahder90}.
The $\gamma_i$ denote the modified Luttinger parameters and can be derived from the
original Luttinger parameters $\gamma_i^L$ by:
\begin{eqnarray}
\label{luttinger}
\nonumber \gamma_c &=& \frac{m_0}{m_e} - \frac{E_\mathrm{P}}{3}\left(\frac{2}{E_\mathrm{g}}+\frac{1}{E_\mathrm{g}+\Delta_{\mathrm{so}}}\right)\\
\nonumber \gamma_1 &=& \gamma_1^\mathrm{L} - \frac{E_\mathrm{P}}{3E_\mathrm{g} + \Delta_{\mathrm{so}}}\\
%\nonumber
\gamma_2 &=& \gamma_2^\mathrm{L} - \frac{1}{2}\frac{E_\mathrm{P}}{3E_\mathrm{g} + \Delta_{\mathrm{so}}}\\
\nonumber \gamma_3 &=& \gamma_3^\mathrm{L} - \frac{1}{2}\frac{E_\mathrm{P}}{3E_\mathrm{g} + \Delta_{\mathrm{so}}}
\end{eqnarray}
$E_\mathrm{cb}$ and $E_\mathrm{vb}$ denote the unstrained conduction and valence band offset, $E_\mathrm{g}=E_\mathrm{cb}-E_\mathrm{vb}$ is
the band gap. $P_0$ is the coupling parameter between conduction and valence bands, $\Delta_{\mathrm{so}}$ denotes
the spin-orbit coupling and
\begin{equation}
E_\mathrm{P}=2m_0\frac{P_0^2}{\hbar^2}
\end{equation}
is the Kane parameter.

\bibliography{paper}
\bibliographystyle{apsrev}

\end{document}